# Modular plasmonic nanopore for opto-thermal gating

*Ali Douaki[1*], Shukun Weng[1], German Lanzavecchia[1,2], Anastasiia Sapunova[1], Annina Stuber[3],*

*Gabriele Nanni[1], Nako Nakatsuka[4], Makusu Tsutsui[5], Kazumichi Yokota[6], Roman Krahne[1], Denis*

*Garoli[1,7*]*

[1] Istituto Italiano di Tecnologia, Via Morego 30, Genova, Italy
[2] Dipartimento di Fisica, Univesità di Genova, Via Dodecaneso 33, Genova, Italy
[3] Laboratory of Biosensors and Bioelectronics (LBB), ETH, Zürich, CH-8092, Switzerland
[4] Laboratory of Chemical Nanotechnology (CHEMINA), Neuro-X Institute, École Polytechnique Fédérale de Lausanne (EPFL), Geneva, CH-1202, Switzerland
[5] The Institute of Scientific and Industrial Research, Osaka University, 8-1 Mihogaoka, Ibaraki, Osaka 5267-0047, Japan
[6] National Institute of Advanced Industrial Science and Technology Takamatsu, Kagawa 761-0395, Japan
[7] Dipartimento di scienze e metodi dell'ingegneria, Università di Modena e Reggio Emilia, Via Amendola 2, 42122, Reggio Emilia, Italy

Email: ali.douaki@iit.it; denis.garoli@unimore.it

# Abstract

Solid-state nanopore gating inspired by biological ion channels is gaining increasing traction due to a large range of applications in biosensing and drug delivery. Integration of stimuli-responsive molecules such as poly(N-isopropylacrylamide) (PNIPAM) inside nanopores can enable temperature-dependent gating, which so far has only been demonstrated using external heaters. In this work, we combine plasmonic resonators inside the nanopore architecture with PNIPAM to enable optical gating of individual or multiple nanopores with micrometer resolution and a switching speed of few milliseconds by thermo-plasmonics. We achieve a temperature change of 40 kelvin per millisecond and demonstrate the efficacy of this method using nanopore ionic conductivity measurements that enables selective activation of individual nanopores in an array. Moreover, the selective gating of specific nanopores in an array can set distinct ionic conductance levels: low, medium, and high (*i.e.*, "0," "1," and "2"), which could be exploited for logical gating with optical signal control. Such selective optical gating in nanopore arrays marks a breakthrough in nanofluidics, as it paves the way towards smart devices that offer multifunctional applications including biosensing, targeted drug delivery, and fluidic mixing.

# Introduction

Solid state nanopores can work as artificial nanofluidic structures that mimic the gating functions of biological ion channels. They are gathering increasing attention because they represent a platform for fundamental research and are highly attractive for diverse applications [1–8]. The ability of nanopores to manipulate the movement of molecules and ions at the nanoscale is a pivotal factor driving this interest. For example, in biosensing applications, controlled gating facilitates the selective detection of biomolecules [9–12]. Similarly, in drug delivery systems, molecular gating enables targeted release of therapeutic agents[13]. Important recent developments in this field involved the incorporation of stimuli-responsive molecules and functional chemical groups into nanopores [14–16]. This advancement has led to the creation of "smart" nanopores and innovative nanodevices, for example by molecular recognition in nanoporous membranes and the development of tunable nanofluidic diodes [17–19]. An interesting material in this respect is poly(N-isopropylacrylamide) (PNIPAM), known for its phase transition at a critical solution temperature (CST) of approximately 32°C [15,20,21]. If used as a functional layer inside a nanopore, PNIPAM can be used as a gate, *i.e.*, below the CST, the polymer expands to block the nanopore, while above this temperature, the polymer collapses to open the pore to ionic flux [22–25]. This temperature-dependent behavior enables thermal modulation of the nanopore channel, offering a straightforward yet effective gating mechanism. However, the gating control reported so far relies on external heaters, which limit the system's functionality, in particular in terms of reaction time in the response to the temperature change [6,18,23,25]. An important remaining challenge is the selective gating of individual nanopores within an array, which requires a precise and highly localized temperature control [6,23]. In this context, optical gating represents an innovative solution that optimizes the nanopore gating by means of thermo-plasmonic effects.

In this work, the challenge of spatially selective gating was addressed by using plasmonic nanostructures for targeted heating with μm spatial resolution. We demonstrate that plasmonic nanopores, functionalized with PNIPAM and surrounded by a plasmonic bullseye structure, can effectively achieve thermoplasmonic gating.

The bullseye was illuminated by a HeNe laser (at 633nm) and focused the electromagnetic field intensity on the aperture of the nanopore, which resulted in rapid and reproducible temperature changes. The effectiveness of this thermoplasmonic modulation is demonstrated through ionic conductivity measurements within single and multiple nanopores. We achieve an On/Off ratio up to 60 of the ionic current and stable switching demonstrated in multiple cycles. The precision, efficiency, and practicality of the thermoplasmonic nanopore gating mark a significant advancement in nanofluidics. This novel approach for selective optical gating of nanopores paves the way to new possibilities in the development of smart nanofluidic systems.

## Results

The thermoplasmonic gating mechanism utilizes a combination of temperature-responsive polymers and targeted laser activation of a plasmonic bullseye structure to control the ion channel conductivity of nanopores. As depicted in Fig. 1, in the absence of laser activation, the PNIPAM layer at the back side of the nanopore remains in a swelled state as the ambient temperature is below the CST. This swelled state of PNIPAM blocks the ion passage through the nanopore, leading to an inactive or 'Off' state. Under laser illumination, the increased electromagnetic field intensity induced by the bullseye leads to a local increase in temperature, and by surpassing the CST, causes the PNIPAM molecules to shrink. This molecular shrinkage opens the pore, allowing the ion flow and switches the nanopore to the 'On' state as previously shown [18,22,23].

We fabricated the nanopores in a Silicon Nitride membrane with a thickness of 100 nm by focused ion beam (FIB) milling. The diameters of these nanopores were determined by two complementary methods: imaging with scanning electron microscopy (SEM), and conductance measurements in a 1 M KCl solution [26]. The latter is based on pores with a conical channel shape and correlated the ionic flux to the effective pore size. We obtain a good agreement between the pore diameters derived from the channel conductance and those imaged *via* SEM (Fig S2), thus providing reliable values for the nanopore sizes that we fabricated. In the next step, PNIPAM was covalently linked to the backside of the pore by self-assembly (as shown in SI – Note 1) to enable the gating functionality. The effective functionalization of nanopores membranes with PNIPAM followed the protocol recently reported by Dahlin *et al.* for plasmonic nanoholes [6]. We note that in the calculations of the effective pore diameter, we assumed a uniform modification of the pore size and membrane surfaces, which takes also the increase in membrane thickness caused by the PNIPAM layer into account.

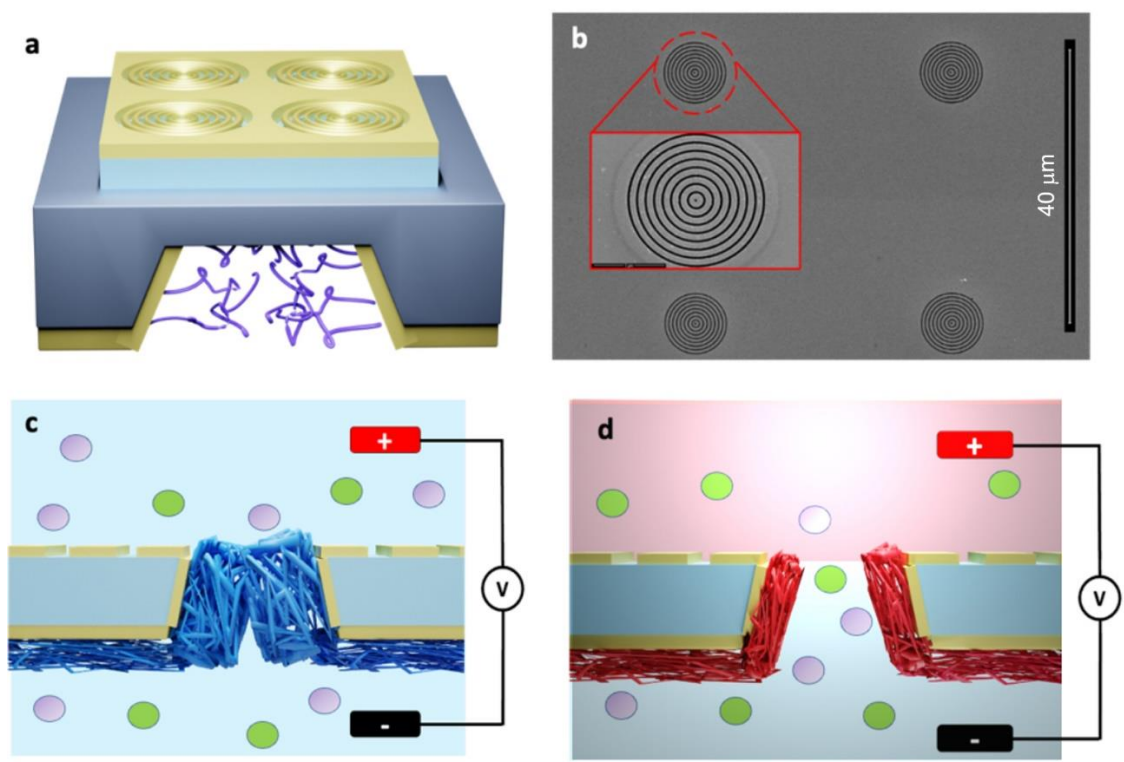

Fig.1: a) illustration of an array of nanopore coated with PNIPAM (bottom side) and bullseye (top side), b) SEM image of an array of the gold plasmonic bullseye used in this work. inset scale bar "5 μm". c) and d) illustration of an individual nanopore coated with PNIPAM (bottom side) and bullseye (top side) with laser off and on, respectively. c) with the laser off the temperature inside the pore was lower than the CST value, hence, the PNIPAM was swelled obstructing the ions flow, d) with the laser "on" the temperature inside the pore was increased above the CST, hence, shrinking the PNIPAM and leading to an increase in the current.

To corroborate the effective immobilization of PNIPAM on the nanopores, X-ray photoelectron spectroscopy (XPS) was performed. The XPS spectra displayed in Fig. 2a show a carbon peak at 283.1 electron volts (eV), a nitrogen peak at 397.9 eV, and an oxygen peak at 529.8 eV that are characteristic of the elemental composition of PNIPAM [27]. The details of the XPS spectra shown in SI, – note#2, provide a detailed chemical characterization of the PNIPAM functionalization. They reveal three types of carbon (C) species, each corresponding to different bonding within the PNIPAM molecule. The peak at 285.0 eV was attributed to carbons in C–C and C–H bonds, commonly found in the backbone of the polymer chain. The peak at 286.1 eV corresponded to carbon nitrogen (C-N) bond that is a key component of the PNIPAM structure. And the peak at 287.8 eV can be associated with C=O bonds, which is indicative of the carbonyl groups in the polymer [28–30]. The distinct presence of these peaks confirmed the successful immobilization of PNIPAM on the nanopores.

We used atomic force microscopy (AFM) to investigate the PNIPAM film morphology on a flat surface (Fig.2b). The knowledge on the film thickness and roughness is important, because the thickness needs to be tailored to the nanopore size to enable switching, and the roughness determines the accuracy of the switching mechanism. The height profile in Fig 2b exhibits a film with ca 10 nm thickness and homogenous surface texture with a roughness of about 1 nm that demonstrates the uniform PNIPAM surface coating.

To confirm the thermo-responsive nature and the hydrophobic/hydrophilic switching behavior of the PNIPAM film, we performed contact angle measurements were at two distinct temperatures above and below CST: 50 °C and 20 °C (Fig. 2c). At 50 °C we observe a large contact angle (84.34 °) that stems from high hydrophobicity. This conformation is attributed to the contraction of polymer chains above the CST of PNIPAM, which is approximately 32 °C. At 20 °C (below CST), the contact angle is much smaller (44.03 °), implying a more hydrophilic and expanded state of the polymer [21,22].

The critical role of PNIPAM thickness variation in the gating mechanism of the nanopore requires detailed control of this parameter for temperatures below and above the CST at which the switching is operated. We used a quartz crystal microbalance with dissipation monitoring (QCM-D) to obtain a qualitative assessment of conformational changes of the PNIPAM film with temperature. This technique is particularly effective in evaluating the restricted movement of PNIPAM when tethered to surfaces, thereby approximating both the dynamics within the nanopore sensors, as well as mimicking a comparable surface chemistry, as a gold coated quartz crystal was used for measurements. A temperature increase from room temperature to above CST resulted in a notable increase of frequency of approximately 60 Hz (Fig S6a), as well as an increased dissipation of around $2 \times 10^{-6}$ (Fig S6) (details in SI – note#3). Based on the measured dissipation limit, we concluded that we can approximate the polymer height based on the Sauerbrey equation, which model the polymer as a rigid film [31]. The Sauerbrey equation relates the change in frequency to the change in mass of the film. Our data yields a decrease in polymer layer thickness of about 10 nm (Fig 2e.) upon increased temperature exposure, which indicates that PNIPAM undergoes a transition to a more compact structure as the temperature surpasses the CST.

The integration of SEM, AFM, XPS, and contact angle analysis provided a comprehensive characterization of gold surfaces grafted with PNIPAM. This multifaceted approach confirmed the successful functionalization of nanopores with PNIPAM. Importantly, these analyses confirmed the ability of PNIPAM to modify its physical state in response to thermal stimuli, notably transitioning from hydrophilic to hydrophobic properties as temperatures exceed its CST. Additionally, AFM and QCM-D analyses elucidated the layer thickness of PNIPAM and the height variation upon heating beyond the CST, leading to insights on the optimal pore diameter for effective pore blockage. Nevertheless, a further examination of the impact of pore size on the current On/Off ratio is reported in Fig. 2f. Below a 30 nm nanopore diameter, the on/off ratio was notably low, attributed to constant pore blockage. This effect was due to the minimal pore size alteration resulting from PNIPAM's thickness change, which prevented pore opening even above the CST. We hypothesized that a high PNIPAM density may restrict the polymer's conformational flexibility and its ability to shrink. Increasing the pore diameter to 30 nm increased the On/Off ratio, indicating that at this diameter, and at temperatures surpassing the CST, PNIPAM could sufficiently undergo conformational changes. This alteration allowed more space for ion passage. Conversely, further enlargement of the pore diameter led to a reduction in the On/Off ratio. In this scenario, even the swollen state of PNIPAM was insufficient to block the pores effectively due to their larger size (as shown in schematic Fig. 2d). In conclusion, these findings highlight the potential of PNIPAM as an effective material for gating nanopores, with an optimal pore diameter of 30 nm for achieving a high on/off ratio.

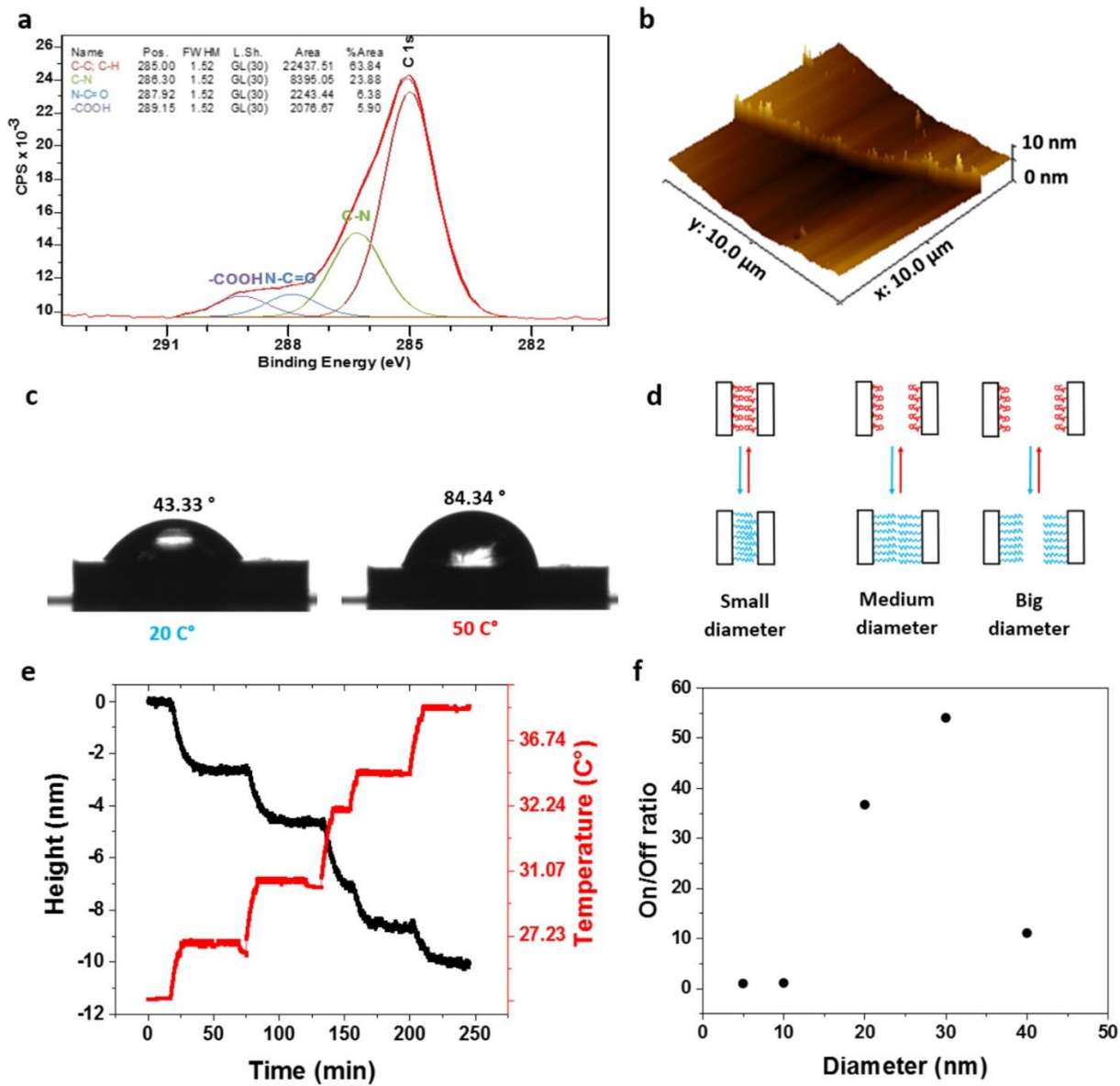

Fig. 2: (a) XPS data of PNIPAM immobilized on a gold layer (representative of the functionalization of the nanopore); b) AFM image of the profile of elongated PNIPAM indicates a layer of 10 nm thick at room temperature; c) contact angle measurement showing the change in hydrophobicity of the PNIPAM film demonstrating the state switch from hydrophilic to hydrophobic; d) schematic illustrations of functionalized nanopores with different diameters; e) Real-time recording of a QCM sensor functionalized with PNIPAM in 100 mM KCl at different temperatures from 23 C° to 40 °C, the PNIPAM-modified sensor exposed to a gradual increase in the temperature showed a decrease in the thickness of the PNIPAM layer, f) current On/Off ratio versus different pore diameters.

To achieve fast and possibly highly localised changes of the PNIPAM conformation we harnessed the nanopores with thermoplasmonic structures. In nanophotonics, plasmonic heating has been successfullly employed through the utilization of the 'bullseye' structure [32]. Figure 1b shows a

SEM micrograph of the bullseye structure that we fabricated with a nanopore in the center (additional images in SI – note#1), and which we illuminated with a laser at 633 nm to induce the localized plasmonic heating. The goal is to increase the local temperature at the nanopore above the CST value of PNIPAM to trigger the morphological and hence the opto/thermo gating of the nanopore. To investigate the temperature changes that can be achieved with this approach, we monitored the conductance change with respect to bullseye illumination, leveraging the temperature-dependent conductivity of potassium chloride (KCl) as electrolyte, for which an increase in temperature is directly proportional to an increase in conductance.

Fig. 3a reports the current-voltage (I-V) curves obtained from a single plasmonic (bullseye) nanopore before the surface functionalization with PNIPAM (bare nanopore). As expected, the conductance of the nanopore depends on the illumination intensity (laser power). We use these measurements to obtain an estimation of the temperature inside the nanopore, and apply COMSOL modeling to numerically calculate the expected temperature inside the nanopore for the different laser excitation powers. Without excitation, the system is at room temperature, while for illumination intensities of 0.5 mW/cm$^2$ and 1 mW/cm$^2$, the simulations show temperatures inside the nanopore of 315 K and 334.7 K, respectively (see Fig. 4). The temperature profile inside the nanopore on the z-axis is shown in Fig. S8. We note that no substantial conductance change was observed in a control experiment illuminating a nanopore without bullseye structure at 1 mW/cm$^2$. This evidences the key role of the thermoplasmonic effect to regulate the temperature at the nanopore with illumination intensities that are sufficiently low to avoid laser damage of the pore of the PNIPAM film.

Fig. 3b and Fig. 3c show the I-V curves in the presence and absence of laser-induced heating, respectively. With the laser off, the internal nanopore temperature was below the CST, with a swelled PNIPAM state that blocks the pore ('Off state'). Under illumination, the nanopore temperature is above CST, and the shrinked PNIPAM conformation results in an opening of the nanopore channel, signifying the 'On state'. Key features of the switching performance are the stability over several cycles, and the time scales at which the temperature and PNIPAM conformational changes are occurring. Figure 3e shows the stability of the on and off currents over several cycles at a fixed bias. The nanopore's electrical response over time for several On and Off switching events is reported in Figure 3d-e. Each switching on event (laser power on) induces a spike in conductivity that decreases exponentially to around 2nA (see Figure S7 for fitting of the decay and evaluation of the decay constants). We attribute this behavior to the capacitive discharging of the charges that accumulated at both sides of the blocked nanopore, and that equilibrate upon pore opening. When the laser is turned off, the current falls to near zero in few ms, evidencing the blockage of the nanopore. To evaluate the response time for the pore opening and closing processes, we analyze the rise and fall times of the current with the 10% and 90% threshold criteria and obtain 3 ms and 4 ms, respectively. We note that these values represent upper limits to the conformational opening and closing process, since they are derived from the current change that is also impacted by the properties of the ionic liquid.

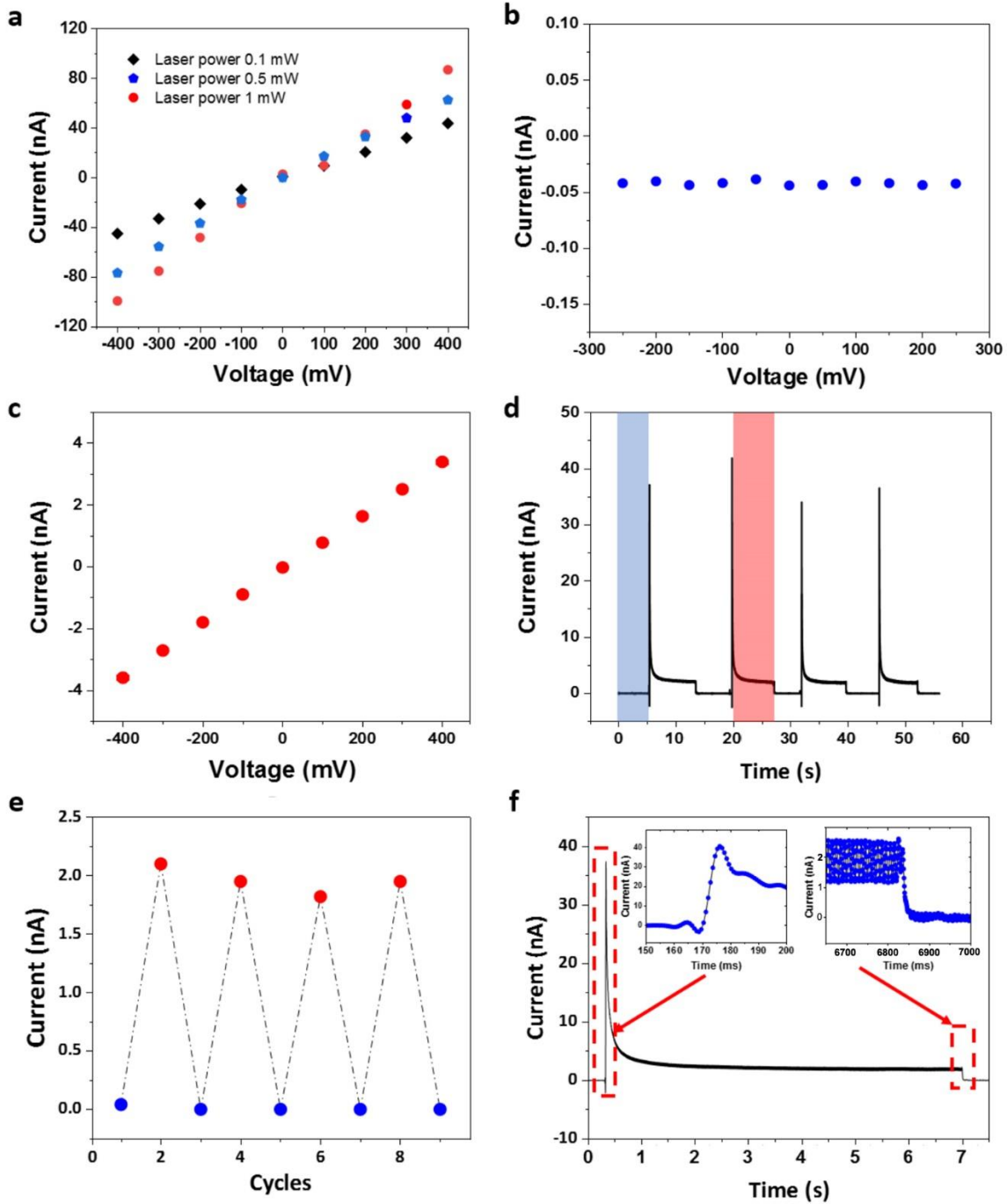

Fig. 3: a) Current–voltage curves of bare nanopore at different laser power 0 mW, 0.5 mW, and 1 mW. (b) Current–voltage characteristics of a nanopore decorated with PNIPAM with the laser in the "off state". C) Current–voltage with of a nanopore decorated with PNIPAM with the laser on "on state" (d) current vs time with cycling where the laser was switched on "red zone" and off "blue zone" at 1V. e) different cycles showing the stability of the pore between switching on and off. f) one cycle enlarged.

The fast thermoplasmonic opening/closing behavior of the nanopores is corroborated by our simulations of the temperature within the nanopore, that rises above the CST in less than 0.5 ms upon illumination (Fig. 4). With switching times of the order of few ms, operation of thermoplasmonic nanopore devices with frequencies up to almost 1 kHz should be possible. We note that the impact of the associated increase and decrease in the local viscosity on the nanopore conductance is anticipated as marginal compared with the giant On/Off ratio observed. [33]

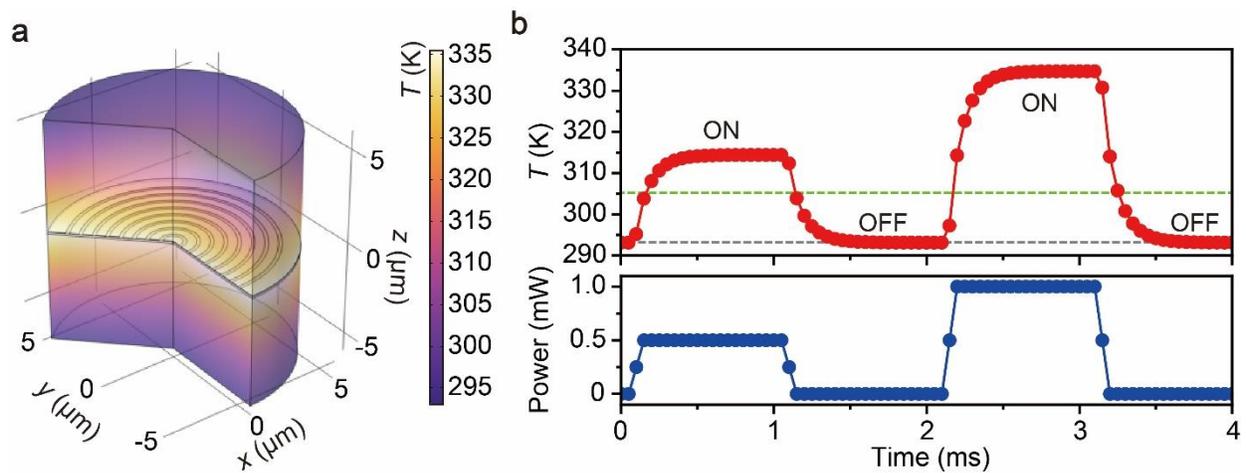

Fig.4: (a) COMSOL modeling of the local temperature in the gold bullseye plasmonic nanopore. Upon illumination with light at 633 nm at 1 mW/cm$^2$. The heat map displays elevated local temperature at steady state around the gold pattern. (b) Temporal response of the temperature at the center of the nanopore, plotted together with the laser power. Green and grey dashed lines denote the threshold temperatures of 32 and 20 °C for switching between the shrinked and extended conformations of the PNIPAM film, respectively.

So far, we discussed the opto/thermal gating of the conductance of a single nanopore. However, our gating concept can be extended to nanopore arrays in which individual pores can be switched separately. In this respect the size of the laser spot (diffraction limited below 1 μm) and the that of bullseye pattern (10 μm diameter in Figure 1b) define the spatial resolution with which individual nanopores can be addressed. Therefore, in principle, nanopores in an array with 12 μm pitch can be controlled on ms timescales, which opens the door to logic gate ionic computing with our thermoplasmonic nanopores. Fig. 5a displays the design of such an array on chip, where each nanopore is surrounded by a gold bullseye pattern that allows for selective and local thermal heating. Multiple pores can be actuated simultaneously by adjusting the laser spot size to illuminate several adjacent pores as in our proof-of-concept demonstration, or via more sophisticated laser beam multiplexing that could allow to address certain sets of nanopores at will. Fast optical addressing of the nanopores could be achieved by galvanic optical mirrors as in commercial confocal scanning microscopes, for example. In the following, we show a proof-of-concept demonstration of the "OR" logic (fig. 5b) gating based on the optical microscope setup in our lab. Here we adjust the laser spot size by focusing and defocusing the microscope optics and address specific chip areas by the movable microscope stage.  Fig. 5 captures the resulting electrical response as multiple nanopores are opto-thermally modulated. The illumination of a nanopore by the laser results in an increase in current. By enlarging the laser spot size, one, then two, three and four pores can be activated simultaneously, which leads to current values of  2 nA, 5 nA, 8 nA, and 10 nA, respectively. Here the conductance of the individual nanopores varies between 2-3 nA due to small differences in nanopore diameter. Addressing of pairs of adjacent nanopores is possible by moving the enlarged laser spot with respect to the pore array.

With this control on the switching multiple nanopores, we can achieve optical gating in the realm of logic gate ionic computing. Building a logic circuit with multiple nanopores on the same chip represents a paradigm shift, as current technology is only able to measure single logic gates independently. Fig. 5b and 5d represent the operational mechanism of the "OR" logic gate using two nanopores (A and B) that can be switched on and off separately or together. The conductance of this nanopore system is reported in Figure 5d. With no laser illumination ([0.0]), both nanopores remain closed, yielding no current flow and representing a binary '0' for both inputs. With the focus of the laser on nanopore A ([1.0]), a distinct increase in the current corresponding to a binary '1' for input A, while the other nanopore (input B) maintains a binary '0'. When the laser targets the second nanopore ([0.1]), input A reverts to '0', and input B transitions to '1'. To obtain the state ([1.1]), the laser illuminates both pores, which leads to larger current as for the individual pores, reflecting a binary '1' for both inputs A and B. This arrangement demonstrates a current-based execution of logical operations, confirming the viability of nanopores array to be orchestrated to perform logical operations.

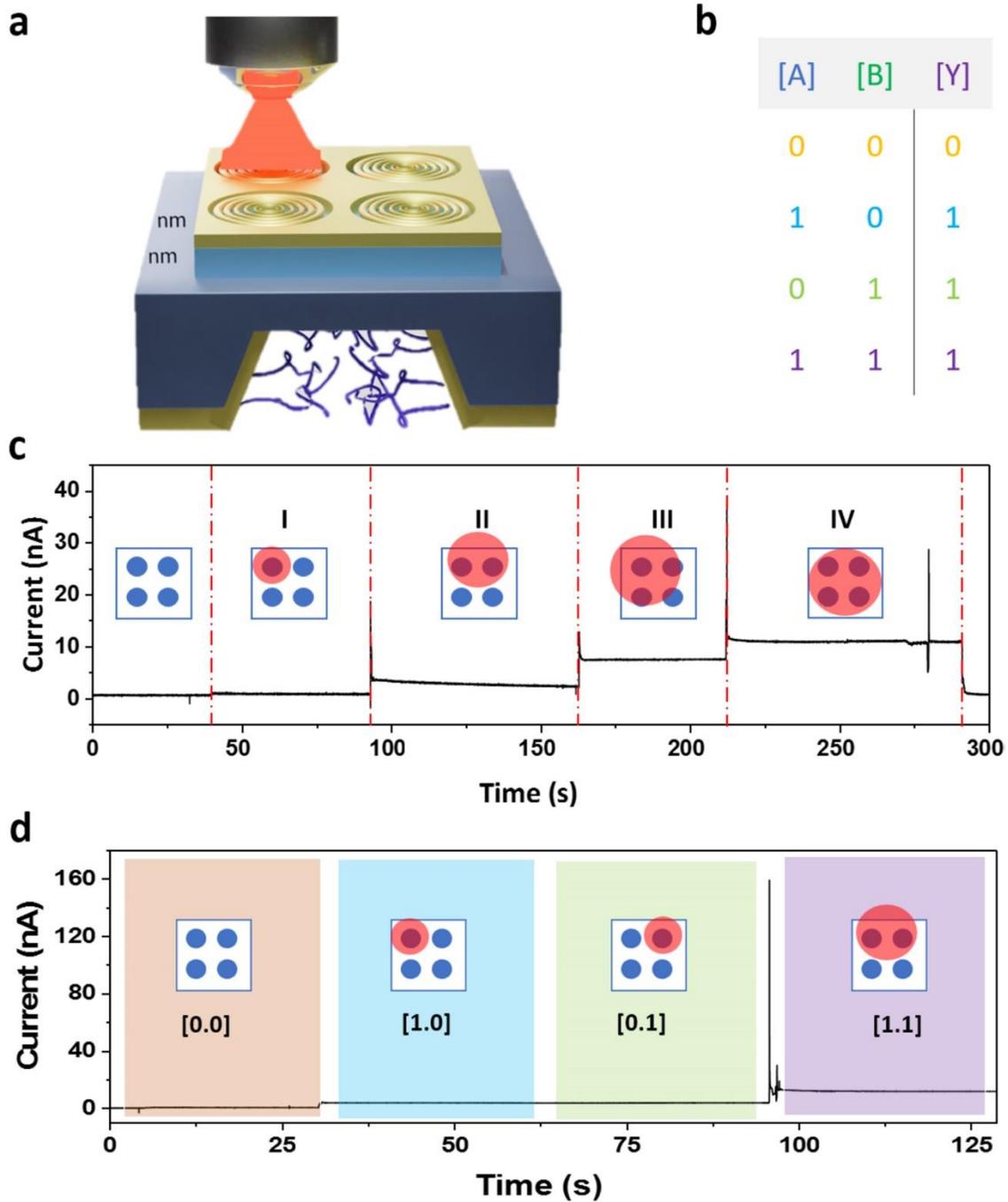

Fig.5: a) illustration of nanopore array, each nanopore is surrounded by bullseye that will lead to a selective heating of nanopores by focusing the laser beam on the desired structure which then leads to a selective opening of nanopores; b) represent the OR logic and the current level shown in panel *d*; c) current vs time it shows an increase in the current due to opening different pores one after another which leads to an increase in the current and then switching them off one after another which led to a decrease in the current, each roman number represents the number of pores opened; d) current from the nanopores array and different following logic gates "OR gate", shows that by modulating two pores with a laser leads to closing nanopore, no current to represent the "off state" of the logic gate then switching on the laser so opening the pore leading to an "on state" logic gate

## Conclusion

In conclusion, this study represents a significant step forward in the field of nanofluidics, illustrating the successful integration of stimuli-responsive molecules, specifically poly(N-isopropylacrylamide) (PNIPAM), into solid-state nanopores alongside a plasmonic structure for opto/thermal gating. This innovative approach overcomes the limitations of traditional temperature-dependent gating methods that rely on external heat sources. By employing thermoplasmonics, this work enabled a selective activation of individual nanopores. This method was validated through ionic conductivity measurements, demonstrating the ability to selectively control nanopores with high precision and On/Off efficiency at a micrometer scale. The high speed in the thermal response (in the millisecond range) enables unprecedented gating efficiency. Furthermore, the introduction of multistage current switching capabilities, exhibiting low, medium, and high (i.e., "0," "1," and "2") quaternary levels of ion conductance, marks a breakthrough in the field. This stable and robust multistage switching could pave the way for multivalued logical gating, managed optically, thereby enhancing the versatility and functionality of nanofluidic devices. This research significantly contributes to the evolution of smart nanofluidic systems, broadening their applicability in areas like biosensing and targeted drug delivery. The ability to control nanopores with such enhanced precision and efficiency holds great promise for future innovations in various scientific and medical fields.

## Methods and materials

**Fabrication Procedures.** Freestanding $Si_3N_4$ membrane chips were prepared following a standard membrane fabrication procedure. In particular, an array of square membranes was prepared on a commercial double-sided 100 nm LPCVD $Si_3N_4$ coated 500 μm Si wafer via UV photolithography, following reactive ion etching and subsequent KOH wet etching. Subsequently, a 3 nm layer of titanium, followed by a 30 nm layer of gold, was evaporated onto both the top and bottom sides of the membranes. This resulted in an overall membrane thickness, inclusive of the adhesion layer, measuring 133 nm. Notably, each silicon chip has dimensions of 5 mm x 5 mm and a thickness of 500 μm. Afterwards, nanopores featuring a diameter of 30 ± 10 nm, and bulls-eye milling (on the top side), were fabricated using a focused Ga$^+$ ion beam. Precise adjustments to milling times were made to achieve the optimized bullseye geometry. Afterwards, 5 nm/20 nm Ti/Au were evaporated using thermal evaporator.

**Grafting PNIPAM.** PNIPAM chains were grafted into the pores of the flat membrane using self-assembled monolayer. The samples were plasma treated for 60 s in oxygen. To create reactive sites on the gold surface for PNIPAM-amine (2500 Mn) attachment, 11-mercaptoundecanoic acid (MUA) was used. A solution of the thiol-based molecule 5mM MUA in ethanol was prepared and the nanopores were immersed for overnight and then the samples were rinsed to remove any unbound molecules and dried under a stream of nitrogen. Finally, the samples were immersed in a solution of 400 mM EDC, 100 mM NHS and 1 wt % PNIPAM-amine in methanol for 3h and then rinsed and dried under nitrogen.

**AFM.** AFM data were taken by non-contact mode AFM using a AFM system XE-100.

**Water Contact Angle Measurement.** Static water contact angle was measured using a Theta Optical Tensiometer (Dataphysics OCAH200) with 5 μL deionized water droplets.

**Quartz Crystal Microbalance.** KSV QCM-Z500 microbalance was used for QCM-D measurements, and PNIPAM was assembled on gold chips. The chips underwent a rigorous cleaning procedure consisting of 2 min sonication in 2-propanol, acetone, and Milli-Q water. The chips were dried using nitrogen and then oxygen plasma cleaned for 10 min. The QCM chip was then sandwiched between two electrodes that apply a voltage to excite the piezoelectric material at its resonance frequency. A flow cell inlet was used to control the temperature of the sensor. The 7th harmonic was used for the analysis of the dissipation, frequency representation and resulting polymer height calculation.

**Electrical Characterization.** Electrical measurements were conducted in 10 mM KCl buffer solution using the same instrument/reader from Elements srl. The measurements were performed either in the dark or under excitation wavelength of 632.8 nm using a Renishaw InVia Raman system with a 50X long distance objective. Prior to the measurement the pores were wetted with DI/IPA for 5 min. Electrical readout of DNA and NPs translocations were conducted in 10 mM KCl by using the same instrument/reader from Elements srl (see details in SI – note#6.

**Numerical Simulations.** Finite element simulations of heat transfer around the gold bullseye plasmonic nanopore were conducted in a two-dimensional cylindrical coordinates system of 6 μm radius and 12 μm height. A model of the 30 nm-thick gold bullseye structure used consisted of 8 rings with a period and width of 460 nm and 180 nm, respectively. At the center, a nanopore of 15 nm radius was defined in a 100 nm-thick $Si_3N_4$ membrane. The thermal conductivity (k), heat capacity at constant pressure (Cp), and density (ρ) of gold were 318 W/(m·K), 129 J/(kg·K), and 19.3 g/cm³ and those of $Si_3N_4$ were 20 W/(m·K), 700 J/(kg·K), and 3.1 g/cm³, respectively. The open space including the nanopore was set to be filled with water of k = 0.59 W/(m·K), Cp = 4190 J/(kg·K), and ρ = 0.998 g/cm³. The heat equation was expressed as:

$$\rho C_p \frac{\partial T}{\partial t} + \nabla \cdot (-k \nabla T) = Q$$

which was solved within a framework of a finite element method. Here, T, t, and Q are the temperature, the time, and the heating rate. The boundary conditions were P = 0.5 mW or 1.0 mW from the gold bullseye and thermal insulation (in normal vector direction) at the edges. All calculations were performed using COMSOL Multiphysics 6.1.

**Acknowledgements**

The authors thank the European Union under the Horizon 2020 Program, FET-Open: DNA-FAIRYLIGHTS, Grant Agreement 964995, the HORIZON-MSCA-DN-2022: DYNAMO, grant

Agreement 101072818 and the HORIZON-Pathfinder-Open: 3D-BRICKS, grant Agreement 101099125.